\documentclass[fleqn,twoside]{article}
\usepackage[headings]{espcrc2}

\readRCS
$Id: espcrc2.tex,v 1.2 2004/02/24 11:22:11 spepping Exp $
\usepackage{graphicx}
\usepackage[figuresright]{rotating}

\newcommand{\AmS}{{\protect\the\textfont2
  A\kern-.1667em\lower.5ex\hbox{M}\kern-.125emS}}

\title{\vspace*{-10mm}
\hfill \hbox{\small{\bf CERN-PH-TH/2009-239}}\vspace*{5mm}\\
Higgs boson decay into heavy quarks and heavy leptons: \\
higher order corrections
\thanks{
delivered at the 3rd Joint International "Hadron Structure - 2009" (HS'09) Workshop, 
                 Tatranska Strba, Slovakia, Aug. 30 -- Sept. 3, 2009;
e-mail: {\em kim@pnpi.spb.ru}. }}

\author{
        Victor T. Kim\address{St. Petersburg  Nuclear Physics Institute of Russian Academy of Sciences
 188300,  Gatchina }}

\runtitle{Higgs boson decay into heavy quarks and heavy leptons: higher-order corrections}
\runauthor{Victor~T.~Kim}

\begin{document}

\begin{abstract}
Theoretical predictions for the decay width
of Standard Model Higgs boson into bottom quarks and $\tau$-leptons,
in the case when $\rm{M_H\leq 2M_W}$, are briefly reviewed.
 The effects of higher order
perturbative QCD (up to $\alpha_s^4$-level)  and QED corrections are considered.
The uncertainties  of the decay width of Higgs boson
into  $\overline{b}b$ and $\tau^+\tau^-$ are  discussed.
\vspace{1pc}
\end{abstract}

\maketitle

\section{Introduction}

Production cross-sections and decay widths of the Standard Electroweak
Model Higgs boson are nowadays among the most extensively analyzed
theoretical quantities (for a recent review, see, e.g.,
\cite{Djouadi:2005gi}, \cite{Assamagan:2004mu}).
Indeed, the  main hope
of scientific community is that this essential ingredient of the Standard
Model may be discovered, if not at Fermilab Tevatron, then at the forthcoming
LHC experiments at CERN.
There is great interest in the ``low-mass'' region
114.5 GeV$\leq \rm{ M_H}\leq 2\rm{M_W}$, because a ``low-mass'' Higgs boson
is heavily favored by Standard Model analysis of the available  precision data.
The lower  bound, 114.5 GeV, was obtained  from the direct searches of Higgs boson at the
LEP2 $e^+e^-$-collider primarily through  Higgs boson decay  into a
$\overline{b}b$-pair.

It should be stressed, that the uncertainties in
$\Gamma(H\rightarrow \overline{b}b)$,   analytically calculated in QCD
using the  $\overline{\rm{MS}}$-scheme at the
$\alpha_s^4$-level  \cite{Baikov:2005rw},
dominate the
theoretical uncertainty for the branching ratio
of $H\rightarrow \gamma\gamma$ decay, which
is considered to be the most important process in
searches for a ``low  mass'' Higgs boson by CMS and ATLAS collaborations at the LHC.

 Here we briefly discuss the  uncertainties  of the QCD predictions for
$\Gamma_{\rm{H\overline{b}b}}=\Gamma(H\rightarrow \overline{b}b)$, including
those which come
from the on-shell mass  parameterizations of this quantity
(previous discussions see in
\cite{Kataev:1992fe}-\cite{Kataev:2008ym} and \cite{Kataev:2009ns})
and  from the resummations of the   $\pi^2$-terms, typical of the
Minkowskian  region (see \cite{Gorishnii:1983cu}-\cite{Bakulev:2008hx}).
We discuss also perturbatiive QED and QCD uncertainties for Higgs boson decay into heavy leptons,
 $\Gamma_{\rm{H\tau\tau}} = \Gamma(H \rightarrow \tau^+\tau^-)$.

\section{QCD corrections for $\Gamma_{\rm{H\overline{b}b}}$ in terms of pole
and running $b$-quark mass}

There are several approaches for Higgs boson decay
$\Gamma_{\rm {H\overline{b}b}}$ in perturbative QCD.
One of them is based on pole (on-shell) mass
consideration \cite{Kataev:1992fe}-\cite{Kataev:1992fe}:

\begin{equation}
\label{OS}
 \Gamma_{\rm{H\bar{b}b}}
  =\Gamma_0^{b}\,
     \bigg[1+\sum_{i\geq 1}
              {\rm \Gamma_i^b}\,
               a_s^i(\rm{M_{H}})
     \bigg]\,,
\end{equation}
where
$\Gamma_0^{b}=(3\sqrt{2}/{8\pi})\rm {G_F M_{H}{m}_b^2}$,
$a_s(\rm{M_{H}}) \equiv \alpha_s(\rm{M_{H}})/\pi$,
$\rm{m}_b$ and $\rm{M_{H}}$ are the pole $b$-quark  and Higgs
boson masses,  and  ${\rm \Gamma_i^b}$-coefficients are $i$th-order
polynomials of large logarithms $L_b=ln(\rm{M_H^2/m_b^2})$.
An another approach is based on $\overline{\rm{MS}}$-scheme framework:
\begin{equation}
\label{MS}
 \Gamma_{\rm{H\bar{b}b}}
  =\Gamma_0^{b}\,
    \frac{{\rm\overline{m}_b^2(M_H)}}
         {\rm{ m_b^2}}\,
     \bigg[1+\sum_{i\geq 1}
              \Delta{\rm \Gamma_i^b}\,
               a_s^i(\rm{M_{H}})
     \bigg]\,,
\end{equation}
where
$a_s(\rm{M_{H}}) \equiv \alpha_s(\rm{M_{H}})/\pi$ and
$\rm\overline{m}_b(M_H)$
are the QCD  running parameters, defined
in the $\overline{\rm{MS}}$-scheme.  The coefficients
$\Delta\rm{\Gamma}_i^b$  can be expressed through
the sum of the following contributions:
\begin{equation}
\Delta\rm{\Gamma}_i^b = \rm{d_i^{E}} + \rm{d_i^{M}} \, .
\end{equation}
Here the positive contributions $\rm{d_i^{E}}$, calculated directly
in the Euclidean region, and $\rm{d_i^{M}}$ are proportional to
$\pi^2$-factors, which are typical for the Minkowski time-like region.

The corresponding expressions for $\Delta\rm{\Gamma}_i^b$
\cite{Gorishnii:1990zu},\cite{Chetyrkin:1996sr}
were derived at
the $\alpha_s^4$-level in Ref. \cite{Baikov:2005rw},\cite{Chetyrkin:1997wm}.
Detailed analysis and results for Higgs decay width
$\Gamma_{\rm {H\overline{b}b}}$ at $\alpha_s^4$-level are presented
in  \cite{Kataev:2008ym},\cite{Kataev:2009ns}), where
the  $\beta$-function of QCD renormalization group (RG) and mass  anomalous
dimension  function $\gamma_m$ \cite{Tarasov:1980au}-\cite{Czakon:2004bu}
were  considered at
the 5-loop level:
\begin{equation}
 \label{beta}
 \frac{da_s}{d\ln\mu^2} =
    \beta(a_s) =
    -\beta_0\,a_s^2
  \dots
    -\beta_4\,a_s^6
    +O(a_s^7)\, ,
\end{equation}
\begin{equation}
 \label{mass}
\frac{d{\ln}\overline{\rm m}_b}
       {d{\ln}\mu^2}   =
   \gamma_m(a_s) =
        -\gamma_0\,a_s
       \dots
       -\gamma_4\,a_s^5
       +O(a_s^6)\,.
\end{equation}
The 5-loop coefficients  $\beta_4$ and $\gamma_4$
are still unknown, but it can be estimated by
Pad\'e approximation procedure, developed in  \cite{Ellis:1997sb}
(see discussion in Refs. \cite{Kataev:2008ym},\cite{Kataev:2009ns}).

It should be stressed, however, that the uncertainties
of the estimated  5-loop contributions to the QCD $\beta$-function
and mass anomalous dimension function $\gamma_m$ are not so important in the
definition of the running of the $b$-quark mass from the
pole mass $\rm{m_b}$ to the pole mass of Higgs boson $\rm{M_H}$. This effect
of running is described by the
solution of the following RG equation:
\begin{eqnarray}
\label{running} {\rm\overline{m}_b^2(M_H)}
 ={\rm\overline{m}_b^2(m_b)}
  \exp\bigg[-2\int_{a_s(\rm{m_{b}})}^{a_s(\rm{M_H})}
\frac{\gamma_m(x)}{\beta(x)}dx\bigg]\\ \nonumber
={\rm \overline{m}_b^2(m_b)}
\bigg (\frac{a_s(\rm{M_{H}})}{a_s(\rm{m_b})}\bigg)^{2\gamma_0/\beta_0}
\bigg(\frac{AD(a_s(\rm{M_ H}))}
{AD(a_s(\rm{m_b}))}\bigg)^2,
\end{eqnarray}
where $AD(a_s)$ is a polynomial  of 4-th order in the QCD expansion
parameter $a_s$
\cite{Kataev:2008ym},\cite{Kataev:2009ns}.

In the Higgs boson mass region of interest, Eq.(\ref{MS})
may be expressed in numerical form as
\begin{eqnarray}
\label{corr}
\frac{\Gamma_{\rm{H\bar{b}b}}}{\Gamma_0^{b}}
&=& \frac{{\rm\overline{m}_b^2(M_H)}}
         {\rm{ m_b^2}}\, \\ \nonumber
&\times& \bigg[1+ 5.667{\it a_s}(\rm{M_H})
+
29.15 {\it a_s}(\rm{M_H})^2 \\ \nonumber
&+&41.76{\it a_s}(\rm{M_H})^3-
825.7{\it a_s}(\rm{M_H})^4\bigg]
\end{eqnarray}
Substituting the value  $a_s(\rm{M_H})\approx 0.0366$ (which corresponds
to $\alpha_s(\rm{M_H}=120~{\rm GeV})\approx 0.115$) into Eq.(\ref{corr}),
and decomposing the coefficients in the Minkowskian series
into  Euclidean  contributions and
Minkowskian-type  $\pi^2$-effects,  one can get from
Ref.\cite{Baikov:2005rw}
the following numbers
\begin{eqnarray}
\label{decomposition}
\frac{\Gamma_{\rm{H\bar{b}b}}}{\Gamma_0^{b}}&
=&\frac{\rm{\overline{m}_b^2(M_H)}}{\rm{m_b^2}}
\bigg[1+0.207+0.039 \\ \nonumber
&+& 0.0020-0.0015\bigg] \\ \nonumber
&=&\frac{\rm{\overline{m}_b^2(M_H)}}{\rm{m_b^2}}
 \bigg[1+0.207
+(0.056-0.017) \\ \nonumber &+& (0.017 -0.015)
+(0.0063-0.0078)~\bigg],
\end{eqnarray}
where the negative numbers  in the round  brackets come from the effects
of analytical continuation. Having a look at Eq.
(\ref{decomposition}) we may conclude that in the Euclidean region
the perturbative series is well-behaved and the
$\pi^2$-contributions typical of the Minkowskian region are also decreasing from order to order.
However, due to the strong   interplay between these
two effects in the third and fourth terms, the latter
ones are becoming numerically comparable.
This feature spoils the convergence
of the perturbation series in the Euclidean region.
Therefore,
to improve the accuracy of the
 perturbative prediction in the Minkowskian region
it seems natural to  sum up these $\pi^2$-terms using
the ideas, developed in the 80s (see, e.g.,
\cite{Yndurain:1980qg}-\cite{Radyushkin:1982kg}).
These ideas now have a more solid theoretical background, see,
 e.g.,  Ref. \cite{Shirkov:2000qv}.

Also, we stress
that  the truncated perturbative expansions
of Eq.(\ref{corr}) have some additional uncertainties. These include
$\rm{M_H}$  and   $t$-quark mass dependent QCD
\cite{Larin:1995sq},
\cite{Chetyrkin:1997vj}  and  QED
\cite{Kataev:1997cq} contributions:
\begin{equation}
\Delta \Gamma_{\rm{H\bar{b}b}}=
\frac{3\sqrt{2}}{8\pi}\rm{G_F}M_{H}\overline{m}_b^2(M_H)\bigg[\Delta^{\rm{t}}
+\Delta^{\rm{QED}}\bigg]
\end{equation}
where  $\Delta^{\rm{t}}$ and $\Delta^{\rm{QED}}$
are defined following  Refs. \cite{Chetyrkin:1997vj}, \cite{Kataev:1997cq}
as
\begin{eqnarray}
\Delta^{\rm{t}}&=&
a_s^2\bigg((3.111-0.667L_t)\\ \nonumber
&+& \frac{\rm{\overline{m}_b^2}}{\rm M_H^2}
(-10+4L_t
+ \frac{4}{3}ln(\rm{\overline{m}_b^2/M_H^2}))\bigg) \\ \nonumber
&+&a_s^3\bigg(50.474-8.167L_t-1.278L_t^2\bigg) \\ \nonumber
&+& a_s^2\frac{\rm{M_H^2}}{\rm{m_t^2}}\bigg(0.241-0.070L_t\bigg) \\ \nonumber
&+&
X_t\bigg(1-4.913 a_s+ a_s^2(-72.117-20.945L_t)\bigg)
\end{eqnarray}
$L_t=ln(\rm{M_H^2/m_t^2})$, $X_t={\rm G_Fm_t^2}/(8\pi^2\sqrt{2})$, $\rm{m_t}$
is the  $t$-quark pole mass, $\rm{\overline{m}_b}=\rm{\overline{m}_b(M_H)}$ and
\begin{equation}
\Delta^{\rm{QED}}=\bigg(0.472-3.336\frac{\rm{\overline{m}_b^2}}{\rm{M_H^2}}\bigg)a
-1.455a^2+1.301aa_s \, .
\end{equation}
Using $a \equiv \alpha(\rm{M_H})/\pi$=0.0027 ( $\alpha(\rm{M_H})^{-1}\approx129$),
$\rm{m_t}=175~{\rm GeV}$, $\rm{M_H}=120~{\rm GeV}$,
$\rm{\overline{m}_b}=2.8~{\rm GeV}$,
$\rm{G_F}=1.1667\times 10^{-5}~{\rm GeV}^{-2}$ we get
\begin{eqnarray}
\label{H1}
\Delta_{\rm{t}}&=&\bigg[4.84\cdot10^{-3}-1.7\cdot10^{-5} \\ \nonumber
\label{H2}
  &+&2.27\cdot 10^{-3}  +1.85\cdot 10^{-4} \\ \nonumber
  \label{t}
&+&3.2\cdot 10^{-3}-5.75\cdot 10^{-4}-2.42 \cdot10^{-4}\bigg] \\
\label{QED}
 \Delta^{\rm QED}&=&\bigg[ 1.1\cdot 10^{-3}-4.5\cdot 10^{-6} \\ \nonumber
 &-&9\cdot 10^{-6}-
1.2\cdot 10^{-4}\bigg].
\end{eqnarray}
Comparing the numbers presented in Eq.(\ref{decomposition}) and
Eq.(\ref{H2})-Eq.(\ref{QED}), we conclude that $\alpha_s^4$-terms
can be neglected at the current level of the experimental precision
of ``Higgs-hunting'' at Fermilab and LHC.
 Indeed, one can see, that even for the light Higgs boson
 the  numerical values of the
order $\alpha_s^4$-contributions to Eq.(\ref{decomposition}) are
comparable with the leading  $\rm{M_H}$- and $\rm{m_t}$- dependent terms in
Eqs. (\ref{H1})-(\ref{t}).

An another approach for $\Gamma_{\rm {H\overline{b}b}}$, where the RG-controllable terms
are summed up, may be written down as
\begin{eqnarray} \label{RG}
 \Gamma_{\rm{H\overline{b}b}}
 &=&
 \Gamma_0^{b}
  \bigg(\frac{a_s(\rm{M_H})}
             {a_s(\rm m_b)}
  \bigg)^{(24/23)}  \\ \nonumber
    &\times&  \frac{AD(a_s(\rm{M_H}))^2}
        {AD(a_s(\rm{m_b}))^2}
       \bigg[1+\sum_{{\rm i}\geq 1}
              \Delta{\rm \Gamma_i^b}\,
               a_s^{\rm i}(\rm M_H)
     \bigg] \\ \nonumber
  & \times &
  \big(1-\frac{8}{3}{\it a_s}(\rm {m_b})
        -18.556\,{\it a_s}(\rm {m_b})^2 \\ \nonumber
        &-& 175.76\,{\it a_s}(\rm{m_b})^3
        -1892\,{\it a_s}(\rm{m_b})^4
  \big)\,,
\end{eqnarray}
where
\begin{eqnarray}
 AD(a_s)^2 &=& 1+2.351\,a_s\\ \nonumber
 &+& 4.383\,a_s^2+3.873\,a_s^3-15.15\,a_s^4.
\end{eqnarray}
Here, an important relation between pole and running masses of Refs.
\cite{Chetyrkin:1999qi},\cite{Melnikov:2000qh},\cite{Kataev:2009ns} has been used.
Detailed comparison of $\rm\Gamma_{H\overline{b}b}$ in RG-improved
(Eq.~(\ref{RG})) and in pole mass truncated (Eq.~(\ref{MS})) approaches
was presented in Refs. \cite{Kataev:2008ym}, \cite{Kataev:2009ns}.

\begin{figure}[b!]
{\includegraphics[width=75mm,height=75mm]{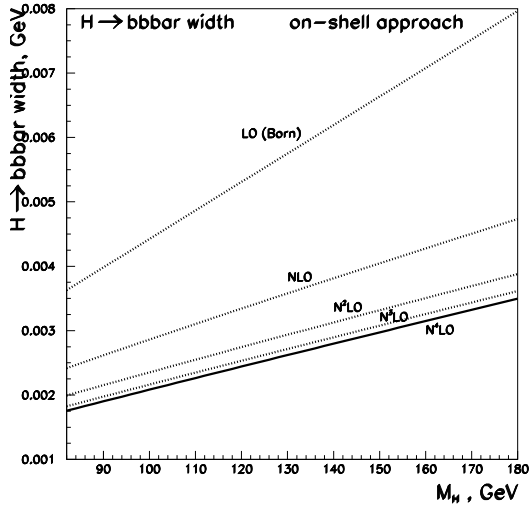}}
\caption{Higgs boson width in the pole (on-shell) mass approach.
\label{fig1}}
\end{figure}

\begin{figure}[t!]
{\includegraphics[width=75mm,height=75mm]{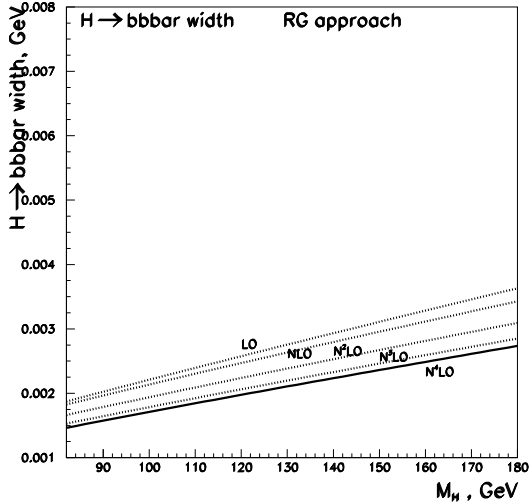}}
  \caption{Higgs boson width in the approach with explicit
RG-resummation.
  \label{fig2}}
\end{figure}

The behavior of the RG-resummed  expressions
for $\rm\Gamma_{H\overline{b}b}$ and  $\rm{R_{H\overline{b}b}}$
are more  stable than  in  the case, when RG-summation of
the mass-dependent terms
is not used \cite{Kataev:1992fe}-\cite{Kataev:2008ym},
\cite{Kataev:2009ns} (Figs. 1,2).
Difference of
$\Delta\Gamma_{\rm H\overline{b}b}$ calculated
the truncated pole-mass approach  and the RG-improved
parametrization of
 $\rm\Gamma_{H\overline{b}b}$ is becoming smaller
 in each successive order of
perturbation theory.

Indeed, for the phenomenologically interesting value
of Higgs boson mass ${\rm M_H}=120~ {\rm GeV}$ we find that
at the  $\alpha_s^2$-level
$\Delta\Gamma_{\rm H\overline{b}b}\approx 0.7~{\rm MeV}$,
while for the  $\alpha_s^3$-level it  becomes smaller, namely
$\Delta\Gamma_{\rm H\overline{b}b}\approx 0.3~{\rm MeV}$.
At the $\alpha_s^3$-level of the RG-improved $\rm\overline{MS}$-scheme
series
one has  $\rm\Gamma_{H\overline{b}b}\approx 1.85 ~{\rm MeV}$
for ${\rm M_H}=120~{\rm GeV}$.
For  this scale the value
of  $\rm\Gamma_{H\overline{b}b}$   with the  explicit dependence from
the pole-mass is {\bf $16~\%$} higher, than its RG-improved estimate.

There are
different approaches  to  the treatment  of  the typical Minkowskian
$\pi^2$-contributions
in the perturbative expressions for physical quantities, which
demonstrated remarkable convergence properties
\cite{Bakulev:2006ex}-\cite{Bakulev:2008hx}. At the moment,
these approaches are developing for different phenomenological applications,
which will alow a comparison with the existing methods.

\section{\bf Higgs boson decay into $\tau^+ \tau^-$}

Width of Higgs boson decay into $\tau^+ , \tau^-$ -leptons in
the $\rm{\overline{MS}}$-scheme
can be read as \cite{Kataev:2009}:
\begin{eqnarray}
\label{MStau}
 \Gamma_{\rm{H{\tau}\tau}}
  &=&\Gamma_0^{\tau}\,
    \frac{{\rm\overline{m}_{\tau}^2(M_H)}}
         {\rm{ m_{\tau}^2}}\,
     \bigg[1+ a(\rm{M_{H}}) \Delta{\rm \Gamma_1^{\tau}}  \\ \nonumber
     &+& a(\rm{M_{H}})^2 \Delta{\rm \Gamma_2^{\tau}}
     + a(\rm{M_{H}})^3 \Delta{\rm \Gamma_3^{\tau}} \\ \nonumber
     &+&
      a(\rm{M_{H}})^2 a_s(\rm{M_{H}}) \Delta^{QEDxQCD}
     \bigg]\,,
\end{eqnarray}
where $\Gamma_0^{\tau}=(\sqrt{2}/{8\pi})\rm {G_F M_{H}{m}_{\tau}^2}$,
$a(\rm{M_{H}}) \equiv \alpha^{\rm{\overline{MS}}}(M_H)/\pi$,
$\rm\overline{m}_{\tau}(M_H)$ are QED running parameters
and $a_s(\rm{M_{H}}) \equiv \alpha_s^{\rm {\overline{MS}}}(M_H)/\pi$ is QCD parameter,
and $\Delta^{\rm QEDxQCD}$ is a mixed QED-QCD correction to the coefficient function.
Evolution of running $\tau$-lepton mass in QED is similar to Eq.~(\ref{running}),
but with $\beta^{\rm QED}$, $\gamma_m^{\rm QED}$, $\Delta{\rm \Gamma_2^{\tau}}$
and $\Delta^{\rm QEDxQCD}$, complicated by quark
fractional electric charge dependence \cite{Kataev:2009}. $\beta^{\rm QED}_3$ is known
since \cite{Gorishnii:1990kd}, and $\gamma^{QED}_3$ \cite{Chetyrkin:1997dh}
is consistent with QED-limit of Ref.~\cite{vanRitbergen:1997va}.
At present for $\Gamma_{\rm{H\tau\tau}}$ to get accuracy of
$\Gamma_{\rm {H\overline{b}b}}$ at  $\alpha_s3$-level it is enough
to keep 2-loop running $\tau$-lepton mass and 1-loop coefficient
function $\Delta{\rm \Gamma_1^{\tau}}$ \cite{Kataev:2009}.

\section{Summary}.

Different approaches based on the  running
and pole $b$-quark masses for the decay
width of the $H\rightarrow\overline{b}b$ process become consistent
in higher orders of perturbative QCD. However, different convergence
in different approaches demonstrates an existence of additional theoretical
QCD uncertainties, which are not usually considered in
phenomenological studies.

Currently, for width of Higgs boson decay into heavy leptons
$\Gamma_{\rm{H{\tau}\tau}}$ to have accuracy of
$\Gamma_{\rm {H\overline{b}b}}$ at  $\alpha_s^3$-level it is enough
to take into account 2-loop running $\tau$-lepton mass and 1-loop coefficient
function $\Delta{\rm \Gamma_1^{\tau}}$.

The author thanks A.~L.~Kataev for fruitful collaboration
on the results presented here.
The author is also grateful to the Local
Organizing Committee of the "Hadron Structure - 2009" (HS09)
Workshop for enjoyable atmosphere and CERN Theory Unit
for their warm hospitality.
The work is supported in parts by Russian Foundation for
Basic Research, grants 08-02-01184à and 06-02-72041-MNTIa,
and the RF President grant NS-378.2008.2.

\end{document}